\documentclass[twocolumn,prb,superscriptaddress,nofootinbib]{revtex4-2}
\usepackage[english]{babel}
\usepackage[ansinew]{inputenc}
\usepackage{boldline,multirow,xcolor,colortbl}
\usepackage{times}
\usepackage{graphicx}
\usepackage{graphics}
\usepackage{amsmath}
\usepackage{amsfonts}
\usepackage{amssymb}
\usepackage{amsbsy}
\usepackage{dsfont}
\usepackage{epstopdf}
\usepackage{makeidx}
\usepackage{subfigure}
\usepackage{color} 
\usepackage{pgf}
\usepackage{bm}
\usepackage{tikz} 
\usepackage[normalem]{ulem}
\usepackage[symbol]{footmisc}
\newcommand{\be}{\begin{equation}}
\newcommand{\ee}{\end{equation}}
\newcommand{\bea}{\begin{eqnarray}}
\newcommand{\eea}{\end{eqnarray}}

\begin{document}
\title{Casimir forces in the flatland: interplay between photo-induced phase transitions and quantum Hall physics}
%Casimir forces in the flatland: impact of topological phase transitions and quantum Hall physics} 
	
% % % % % % % % % % % % % % % % % % % % % %

\author{Y. Muniz}
\email{yurimuniz@pos.if.ufrj.br}
\affiliation{Instituto de F\'{\i}sica, Universidade Federal do Rio de Janeiro, Caixa Postal 68528, Rio de Janeiro 21941-972, RJ, Brazil}

\author{C. Farina}
\email{farina@if.ufrj.br}
\affiliation{Instituto de F\'{\i}sica, Universidade Federal do Rio de Janeiro, Caixa Postal 68528, Rio de Janeiro 21941-972, RJ, Brazil}

\author{W. J. M. Kort-Kamp}
\email{kortkamp@lanl.gov}
\affiliation{Theoretical Division, Los Alamos National Laboratory, MS B262, Los Alamos, New Mexico 87545, United States}	

\begin{abstract}
We investigate how photo-induced topological phase transitions and the magnetic-field-induced quantum Hall effect simultaneously influence the Casimir force between 
two parallel sheets of staggered two-dimensional (2D) materials of the graphene family. We show that the interplay between these two effects enables on-demand switching 
of the force between attractive and repulsive regimes while keeping its quantized characteristics. 
We also show that doping these 2D materials below their first Landau level allows one to probe the 
photoinduced topology in the Casimir force without the difficulties imposed by a circularly polarized laser. We demonstrate that the magnetic field has a huge impact on the thermal Casimir effect 
for dissipationless materials, where the quantized aspect of the energy levels leads to a strong repulsion that could be measured even at room temperature.

\end{abstract}

\maketitle

%\section{Introduction}

The Casimir effect is an interaction induced by quantum vacuum fluctuations of the electromagnetic field and exists between material 
bodies \cite{bordag2009,dalvit2011,buhman2013,bordag2001,woods2016,farina2006}. It was originally proposed by H.B.G. Casimir in 1948 as an attractive force 
between two neutral perfectly conducting plates \cite{casimir1948}. Fundamentally, it is a dispersive force between two bodies at length scales where 
electromagnetic retardation becomes relevant.  The ever-present dispersive interactions are one of the most intriguing kind of intermolecular forces and
play an important role not only in different areas of physics, but also in chemistry, biology, and engineering \cite{parsegian2006}. For instance, dispersive forces are
crucial to understand the stability of colloids \cite{israelashivili2011}, the drug binding in proteins and the doublehelix structure of DNA \cite{distasio2012}, and even the adhesion of geckos to walls \cite{autumn2000,autumn2002}.
In the manufacturing and operation of devices at the micro- and nanoscales these forces also have severe consequences since they dominate at very small distances and may cause stiction due 
to their tipically attractive nature \cite{serry1998,buks2001,broer2013,sedighi2015,rosa2018}. Therefore, it is very important to study dispersive interactions and master their manipulation not only in magnitude but also in sign, i.e., to tune the
attractive or repulsive character of the force. In this direction, real materials have shown to support a wealth of different dispersive 
interactions \cite{mostepanenko2009,woods2016}, and the strong development of materials science over the last few decades has given a new impetus to the study of the 
Casimir force and its possible control \cite{johnson2011,tang2017}. In particular, switching between repulsive and attractive Casimir forces using external agents 
can be realized with topological insulator plates \cite{grushin2011,cortijo2011,lopez2014,wilson2015}, and superparamagnetic metametamaterials \cite{ma2014}.

The reduced dimensionality of graphene and its related nanostructures has also led to the discover of
novel behaviours in the Casimir interaction, which have a strong dependence on temperature and 
doping \cite{santos2009,woods2010,sarabadani2011,bordag2012,mostepanenko2013,tsoi2014,gobre2014,bimonte2017,mostepanenko2020}. Meanwhile, quantum Hall physics arising 
from externally applied magnetic fields result in a complex and rich magneto-optical response of graphene \cite{gusynin2005,gusynin2007,goerbig2011}. 
This was already exploited for tuning and screening 
dispersive interactions, which were shown to be quantized and significantly weaker when compared to the situation without the field \cite{macdonald2012,tarik2014}. On the other hand, 
two-dimensional (2D) Dirac materials such as silicene, germanene, stanene, and plumbene are representatives of the graphene family and present several possible 
topological phase transitions under external fields due to their significant spin-orbit coupling and finite staggering \cite{ni2012,nicol2012,ezawa2013,ezawa2015,kortkamp2017}. 
These phase transitions strongly impact their Casimir forces, in which different scaling laws, repulsion, and force quantization are possible \cite{woods2017}. Nevertheless, 
the effects that can be brought to the Casimir force as a consequence of coexistent topological and quantum Hall features are yet to be unveiled.

Here we demonstrate that the introduction of a magnetic field is very convenient for switching the sign of the Casimir force between graphene family materials 
while keeping its quantized behaviour. 
We show that the zero-temperature long distance Casimir force between two identical parallel sheets is proportional to the product of their corresponding Chern numbers for every non-trivial topological state, 
and can be made attractive or repulsive by properly doping each plate. 
Furthermore, all photo-induced topologies are mimicked by tuning the chemical potential and do not require energy absorption from a circulary polarized laser, 
in contrast to what is feasible without the magnetic field. We also investigate the Casimir effect for finite temperatures and show that no matter how we model the materials with or without 
dissipation, a huge change in their Casimir energies may be achieved. For a dissipative system, the Casimir force at high temperatures is attractive and does not depend on any material 
properties, while for non-dissipative monolayers we observe repulsion in a wide range of temperatures. Due to their direct parallel to the Drude and plasma models for metals, 
this contrasting behaviour of the Casimir interaction in both situations touches fundamental questions underlying Casimir physics, and could be useful in future investigations.

This paper is organized as follows. In Section I we describe the framework in which we carried our calculations of the Casimir energy. We also give a brief overview on the graphene family 
materials main properties and magneto-optical response. In Section II we present our results on the zero-temperature Casimir interaction. We then discuss the long-distance behaviour of 
the Casimir energy in the phase space, its distance dependence, and how changing physical quantities such as magnetic field, chemical potential or the dissipation affects the force. In 
section III we discuss the thermal effects on the Casimir energy for dissipative and non-dissipative monolayers. Section IV is left for final remarks and conclusions.
\section{Model and methods}
\begin{figure*}
\begin{center}
\includegraphics[width=0.9\linewidth,keepaspectratio]{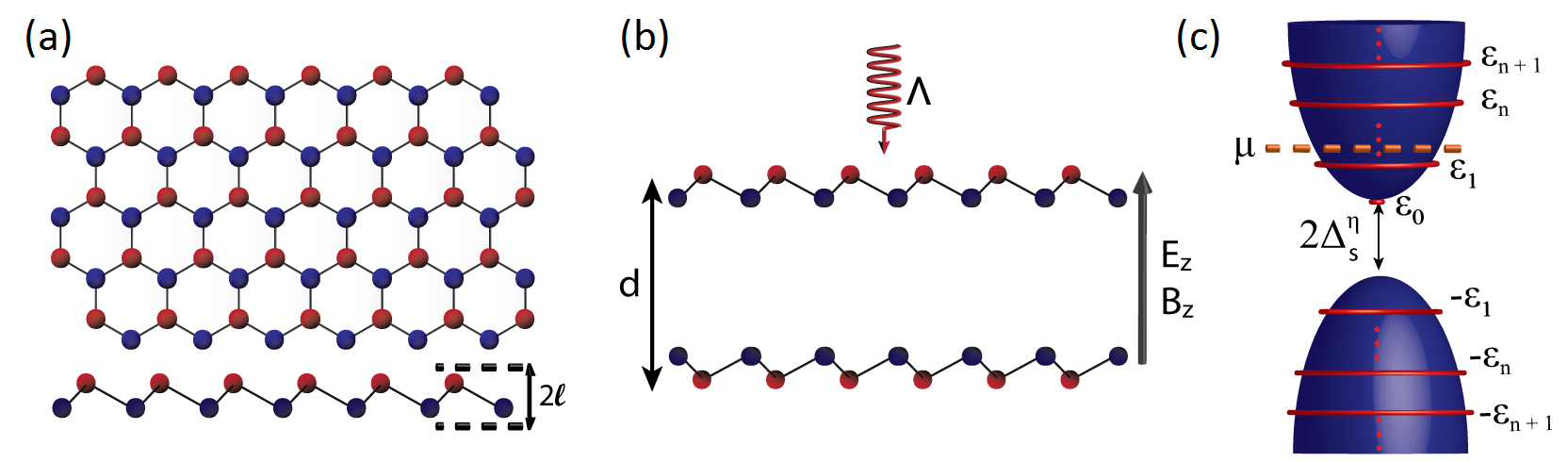}
\caption{{(a)} Top and side views of the graphene family honeycomb structure. The red and blue colored atoms belong to inequivalent sublattices with a finite staggering $2\ell$ between them. {(b)} Schematics of the system under study. Two graphene family monolayers separated by a distance $d$ and subjected to externally applied electric ($E_z$) and magnetic ($B_z$) fields, and a circulary polarized laser ($\Lambda$); {\bf c)} Low-energy band structure around a given $K$ or $K'$ point. For a non-zero magnetic field, the quantized Landau levels (red circles) are built on top of the Dirac cone.}
\label{fig1}
\end{center}
\end{figure*}
In this section, we focus our attention on the Casimir energy per unit area (denoted by $E$) between two layers of the graphene family materials. Like graphene, these materials have a honeycomb structure, however, the two atoms in the unit cell are arranged in staggered layers separated by a distance $2\ell$, as depicted in Fig. 1a \cite{ezawa2015}. The schematics of the system under study is shown in Fig. 1b, where the layers are separated by a distance $d$ and subjected to a circulary polarized laser field of frequency $\omega_0$ and intensity $I_0$, and an out-of-plane static and uniform electric ($E_z$) and magnetic ($B_z$) fields. The Casimir energy can be obtained by the Lifshitz formula, which is valid for $d$ much larger than the interatomic distances of the materials, and is given by \cite{woods2016,reynaud2006}
\begin{equation}
{E} = k_BT \sideset{}{'}\sum_n\int\frac{d^2{\bf k}_\parallel}{(2\pi)^2}\mbox{log det}\left(1 - \mathbb{R}_1\cdot\mathbb{R}_2e^{-2k_{z,n}d}\right), \label{Lifshitz}
\end{equation}
where $k_{z,n} = \sqrt{{\bf k}_\parallel^2 + \xi^2_n/c^2}$, $\xi_n = 2\pi n k_BT/\hbar$ with $n = 0, 1, ...$ are the Matsubara frequencies, and the prime on the summation symbol indicates that the term $n = 0$ must be multiplied by $1/2$. Also, $\mathbb{R}_i$ are the reflection matrices for each sheet, with rows given by the Fresnel coefficients for incident s- and p- polarized waves, which are given by \cite{woods2017} 
\begin{align}
\mathbb{R}_{ss} &= -\frac{2\pi}{\delta_n}\left[ \frac{\sigma_{xx}\xi_n}{k_{z,n}c^2} + \frac{2\pi}{c^2}\left(\sigma_{xx}^2 + \sigma_{xy}^2\right) \right], \label{Rss}
\\
\mathbb{R}_{sp} &= \mathbb{R}_{ps} = \frac{2\pi\sigma_{xy}}{\delta_n c},\label{Rsp}
\\
\mathbb{R}_{pp} &= \frac{2\pi}{\delta_n}\left[ \frac{\sigma_{xx}k_{z,n}}{\xi_n} + \frac{2\pi}{c^2}\left(\sigma_{xx}^2 + \sigma_{xy}^2\right) \right],\label{Rpp}
\end{align}
where $\delta_n = 1 + \frac{2\pi\sigma_{xx}}{c}\left(\frac{\xi_n}{k_{z,n}} + \frac{k_{z,n}}{\xi_n}\right) + \frac{4\pi^2}{c^2}\left(\sigma_{xx}^2 + \sigma_{xy}^2\right)$ and the conductivity components are evaluated at the imaginary Matsubara frequencies $i\xi_n$. It is important to notice that since the reflections occur inside the cavity, we must perform the transformation $\sigma_{xy} \rightarrow -\sigma_{xy}$ for one of the monolayers.

The conductivity components for graphene family materials are well known and can be cast as \cite{nicol2013,dalvit2018}, 
\begin{align}
\sigma_{xx} &= \sigma_{yy} = \frac{iE_B^2\sigma_0}{\pi}\sum_{\eta,s}\sum_{n,m}\frac{f_m - f_n}{\varepsilon_n - \varepsilon_m}\cr
&\times \frac{(A_m^+A_n^-)^2\delta_{|n|,|m| - \tilde{\eta}} + (A_m^-A_n^+)^2\delta_{|n|,|m| + \tilde{\eta}}}{\varepsilon_m - \varepsilon_n + i\hbar(\xi + \Gamma)}, \label{sigmaxx}
\\
\sigma_{xy} &= -\sigma_{yx} = -\frac{E_B^2\sigma_0}{\pi}\sum_{\eta,s}\sum_{n,m}\eta\frac{f_m - f_n}{\varepsilon_n - \varepsilon_m}\cr
&\times \frac{(A_m^+A_n^-)^2\delta_{|n|,|m| - \tilde{\eta}} - (A_m^-A_n^+)^2\delta_{|n|,|m| + \tilde{\eta}}}{\varepsilon_m - \varepsilon_n + i\hbar(\xi + \Gamma)}. \label{sigmaxy}
\end{align}
Here, $\varepsilon_n = \mbox{sgn}(n)\sqrt{|n|E_B^2 + (\Delta_s^\eta)^2}$ for $n \neq 0$ and $\varepsilon_0 =-\tilde{\eta}\Delta_s^\eta$ are the quantized eigenenergies of the system, with $\tilde{\eta} = \mbox{sgn}(eB)\eta$, $E_B = \sqrt{2v_F^2\hbar|eB_z|}$ being the relativistic analog of the cyclotron energy, $v_F$ is the Fermi velocity, and $\Delta_s^\eta$ is half the mass gap. The Dirac mass naturally arises from the spin orbit coupling (SOC) $\lambda_{SO}$, but is modified by the applied electric field and circularly polarized laser, leading to  $\Delta_s^\eta = -\eta s \lambda_{SO} + e\ell E_z + \eta\Lambda$, where $\Lambda = \pm 8\pi\alpha v_F^2I_0/\omega_0^3$ and $\alpha \approx 1/137$ is the fine structure constant. The SOC for silicene, germanene, stanene, and plumbene are $\lambda_{SO} \approx 3.9,43,100,200$ meV \cite{ezawa2015,xiang2017}, respectively. Also, $A_n^\pm = \sqrt{|\varepsilon_n| \pm \mbox{sgn}(n)\Delta_s^\eta/2|\varepsilon_n|}$ for $n \neq 0$, $A_0^\pm = (1 \mp \tilde{\eta})/2$, $\sigma_0 = \alpha c/4$ is graphene's universal conductivity. $f_n = 1/[e^{(\varepsilon_n - \mu)/k_BT} + 1]$ denotes the Fermi Dirac distribution, $\mu$ is the chemical potential, $T$ is the absolute temperature, and $\Gamma$ is the dissipation rate. The summation indices $\eta$ and $s$ are the valley and spin degrees of freedom, respectively, and only admit values of $\pm 1$. The Landau energy levels are illustrated in Fig. 1c for a single Dirac cone. Terms originating from Rashba physics are not considered due to their small contribution \cite{ezawa2013}. We finally notice that the conductivity tensor is a real function of $\xi$. 
\section{Zero-temperature Casimir energy}
In all subsequent discussions, we shall always normalize the Casimir energy between graphene family materials by the zero-temperature neutral graphene/graphene Casimir interaction in the long distance regime, namely, \cite{woods2010}
\begin{equation}
E_g = -\frac{\hbar c \alpha}{32\pi d^3}. \label{GrapheneCasimir}
\end{equation}
Since the Casimir force per unit area is given by $F = -\partial E/\partial d$, we have $F_g < 0$, which results in an attraction between the 
plates. We start our analysis by investigating the Casimir energy in the long distance regime. For $T = 0$, the Casimir energy is given by an integral 
over imaginary frequencies $i\xi$, which can be obtained by the transformation $k_BT \sideset{}{'}\sum_n \rightarrow \frac{\hbar c}{2\pi}\int_0^\infty d\xi$. Since the integrand inside the Lifshitz formula decays with $e^{-2\xi d/c}$, for large separations ($d >> \hbar c/\lambda_{SO}$) we expect the graphene family low-frequency optical response to be dominant. Therefore, a good approximation is to expand the integrand as a power series of $i\xi$ and retain only the leading contribution (while keeping the exponential factor). Since the reflection matrices depend on frequency through the conductivity tensor, we shall first expand $\sigma_{xx}$ and $\sigma_{xy}$. In the absence of dissipation, eqs. \eqref{sigmaxx} and \eqref{sigmaxy} show that the longitudinal conductivity is an odd function of frequency, while the Hall conductivity is an even function of $\xi$. Hence, $\sigma_{xx}(i\xi) = \sigma_{xx}'(0)\xi + O(\xi^3)$ and $\sigma_{xy}(i\xi) = \sigma_{xy}(0) + O(\xi^2)$. For $k_BT = \hbar\Gamma = 0$, one can show that the DC Hall conductivity presents clear signatures of topology and quantum Hall effect. It is given by $\sigma_{xy} = 2\sigma_0(C_{\text{ph}} + C_{\text{QH}})/\pi$, where $ C_{\text{ph}}$ is the Chern number associated with photoinduced topology, and $C_{\text{QH}}$ is the Chern number associated with the quantum Hall effect, namely \cite{dalvit2018}
\begin{align}
C_{\text{ph}} &= -\frac{1}{2}\sum_{\eta,s}\theta\left(\varepsilon_1 - |\mu|\right)\eta\,\mbox{sgn}\left(\Delta_s^\eta + \tilde{\eta}\mu\right),
\\
C_{\text{QH}} &= -\mbox{sgn}(eB\mu)\sum_{\eta,s}\theta\left(|\mu| - \varepsilon_1\right)\left(N_s^\eta + 1/2\right),
\end{align}
with $N_s^\eta$ being the number of filled Landau levels per Dirac cone. By keeping only the DC Hall conductivity we have $\mathbb{R}_{sp} = \mathbb{R}_{ps} \approx \alpha (C_{\text{ph}} + C_{\text{QH}})$ and $\mathbb{R}_{ss} = \mathbb{R}_{pp} \approx 0$, where we discarded $\sigma_{xy}^2$ since it gives a higher order contribution in $\alpha$. By applying the relation $\mbox{log det}\left(1 + A\right) = \mbox{Tr log}\left(1 + A\right) \approx \mbox{Tr}(A)$ to the integrand of the Lifshitz formula, we obtain
\begin{equation}
\frac{E^{(0)}}{E_g} = -\frac{4\alpha}{\pi}(C_{1,\text{ph}} + C_{1,\text{QH}})(C_{2,\text{ph}} + C_{2,\text{QH}}). \label{LongDistanceCasimir}
\end{equation}
Since $E^{(0)}$ has the same distance dependence as $E_g$, the Casimir force normalized by the graphene-graphene force ($-\partial E_g/\partial d$) is also 
given by the previous equation. As a consequence, a negative (positive) normalized energy implies a repulsive (attractive) force. Hence, for two identical 
layers in the same non-trivial topological phase the Casimir force in the long distance regime is repulsive and proportional to the square of the full 
Chern number ($C = C_{\text{ph}} + C_{\text{QH}}$) characterizing the corresponding phase of the materials. However, the Casimir force is 
attractive if the materials are in different topological phases, which in this configuration could only be achieved by doping the materials at different levels 
since the external fields applied to both monolayers are the same. 
We mention that the interplay between photo-induced topology and quantum Hall physics requires $E_B \sim \lambda_{SO}$, which can be accessed in the 
graphene family for magnetic fields less than or of the order of a few teslas. 

When one layer is in a trivial topological phase ($C = 0$), which can be achieved for external fields such that $ C_{\text{ph}} = - C_{\text{QH}}$ or $C_{\text{ph}} = C_{\text{QH}} = 0$ (the latter only being possible if $|\mu| < \varepsilon_1$ for every Dirac cone),  the first contribution to the Casimir energy arises from the longitudinal conductivity. In the situation where $C_1 = C_2 = 0$, the reflection matrices are given by $\mathbb{R}_{sp} = \mathbb{R}_{ps} \approx 0$, $\mathbb{R}_{ss} \approx -\frac{2\pi\sigma_{xx}'\xi^2}{c^2k_z}$, and $\mathbb{R}_{pp} \approx 2\pi\sigma_{xx}'k_z$, where $\sigma_{ij}' = \partial\sigma_{ij}/\partial\xi\vert_{\xi \rightarrow 0}$, hence
\begin{equation}
\frac{E^{(0)}}{E_g} \approx \frac{144\pi}{5\alpha d^2}\sigma_{1,xx}'\sigma_{2,xx}'.
\end{equation}
In the above formula, it is important to notice that, since the conductivity is proportional to the fine structure constant, $E/E_g$ is linear in $\alpha$, such as the case where $C_1,C_2 \neq 0$. However, the normalized energy goes with $1/d^2$, decaying much faster with the distance. If only $C_1 = 0$ but $C_2 \neq 0$ (for instance, by considering monolayers of two different graphene family materials), we must also consider the terms proportional to $\sigma_{2,xy}^2$ inside the reflection matrices in order to obtain the first non zero contribution. By doing that, we obtain
\begin{equation}
\frac{E^{(0)}}{E_g} \approx \frac{8\alpha}{d}\sigma_{1,xx}'C_2^2.
\end{equation}
This time, the normalized energy is proportional to $\alpha^2/d$, which is also much smaller than the situation where the Chern numbers are finite. In summary, the message from the previous equations is that if $C_1$ or $C_2$ vanish, the long distance Casimir force is attractive, smaller in magnitude, and decreases faster with the distance than the case of non-trivial topological states. Therefore, eq \eqref{LongDistanceCasimir} is the dominant approximation to the Casimir energy in the long distance regime, in contrast to what we get in the abscence of a magnetic field, where the Casimir force near the boundaries of the phase diagram is attractive and much higher than far from them\cite{woods2017}.

\begin{figure}
    \begin{center}
        \includegraphics[width=0.98\linewidth]{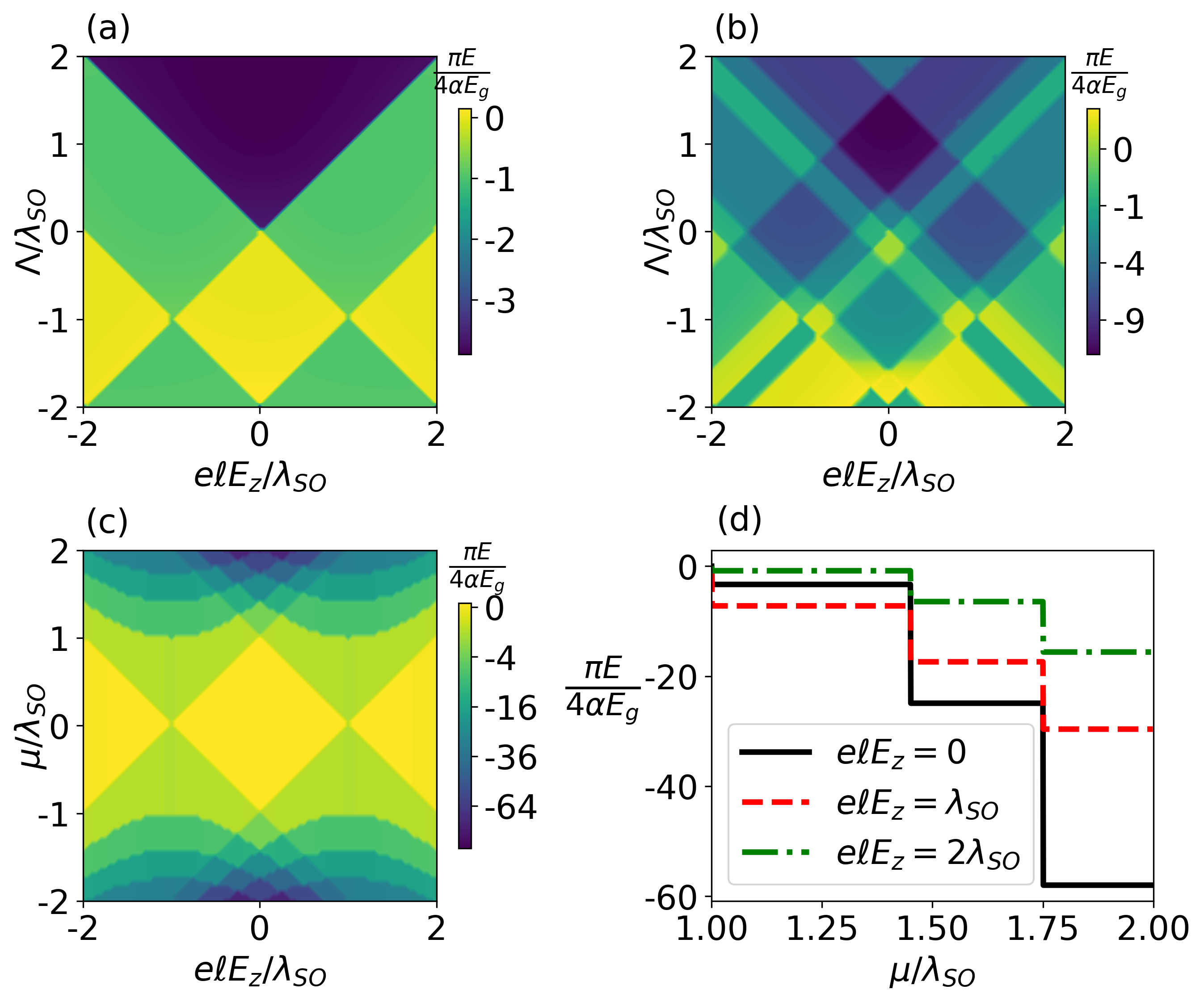}(
        \caption{Casimir energy in the $(E_z,\Lambda)$ plane for (a)  $\mu = \lambda_{SO}$ and $E_B = 1.2\lambda_{SO}$ and  (b) $E_B = 0.8\lambda_{SO}$. {(c)} Normalized Casimir energy in the $(E_z, \mu)$ plane for $\Lambda = 0$ and $E_B = \lambda_{SO}$. { (d)} Normalized Casimir energy as a function of the chemical potential for $\Lambda = 0$. In all plots the distance between the plates is given by $d\lambda_{SO}/\hbar c = 10$ and the dissipation is set to zero.}
        \label{fig2}
    \end{center}
\end{figure}

In Fig. 2a we plot the Casimir energy for two identical layers separated by $d\lambda_{SO}/\hbar c = 10$ in the $(E_z,\Lambda)$ plane for 
$\mu_1 = \mu_2 = \lambda_{SO}$ (denoted by $\mu$) 
and $E_B = 1.2\lambda_{SO}$. In all regions where $C \neq 0$ the normalized Casimir energy is negative and approximately given by eq. 
\eqref{LongDistanceCasimir}, presenting clear signatures of topology. When $C = 0$, the right-hand side of eq. \eqref{LongDistanceCasimir} vanishes, 
but we get a small attractive force between the plates, which arises from further expansions of the reflection matrices in the Lifshitz equation. 
Since $|\mu| < \varepsilon_1$ for every cone (which is achieved in our plot with the stronger condition $|\mu| < E_B$), we have $C_\text{QH} = 0$, 
implying that we are only able to probe the photo-induced topology. 
This feature has its roots in the quantum anomaly of the zeroth Landau level \cite{karch2010}, which has a twice smaller degeneracy and its energy does not depend on the magnetic field. 
As a consequence, the Chern number depends only on $\Delta^\eta_s\left[\Lambda\right] + \tilde{\eta}\mu = \Delta^\eta_s\left[\Lambda + \mbox{sgn}(eB)\mu\right]$, and thus 
changing the chemical potential induces the same effect in the phase diagram as modifying the circularly polarized laser intensity. Furthermore, the diagram 
is also symmetric with respect to $E_z = 0$ and antisymmetric with respect to $\Lambda = -\mbox{sgn}(eB)\mu$. In the situation where $\mu = 0$, the Chern number is given 
by $C = -\sum_{\eta,s}\eta\,\mbox{sgn}\left(\Delta_s^\eta\right)/2$, being the same as in the case without the magnetic field \cite{ezawa2013}. 
For $|\mu| > E_B$, the first Landau level may be occupied in a given cone depending on the values of $E_z$ and $\Lambda$, resulting in 
$C_\text{QH} \neq 0$ and a wealth of quantum Hall topological phases (Fig. 2b). Fig 2c is a plot of the normalized Casimir energy in the $(E_z, \mu)$ 
space for $\Lambda = 0$.  The shape of the diagram in the region $|\mu| < E_B$ resembles Fig. 2a due to the similar role played by $\mu$ and $\Lambda$ under this condition.

This shows that the chemical potential can be used as a substitute of the circularly polarized laser field to probe the photo-induced topological features of the 
Casimir effect. Furthermore, the chemical potential presents an advantage over the laser field since it does not cause an increase of the system's temperature. 
For highly doped materials ($|\mu| > E_B$), the quantum Hall effect dominates the Casimir interaction and hyperbola-like curves define the boundaries between different 
topological phases. In Fig 2d we display the normalized Casimir energy as a function of the chemical potential. When $\mu$ crosses one of the Landau levels in a single 
Dirac cone from below the Casimir interaction increases, becoming more repulsive. This same behaviour is found in the graphene-graphene interaction in the presence of 
an external magnetic field \cite{macdonald2012}. However, here the position and height of the quantum Hall jumps can be tuned with an externally applied electric field, 
which can move the energy levels upwards or downwards.

\begin{figure}
    \begin{center}
        \includegraphics[width=0.98\linewidth]{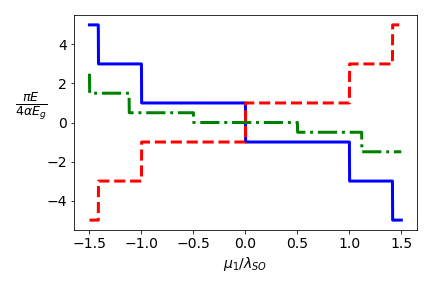}(
        \caption{Casimir energy as a function of the chemical potential of a single monolayer for $\{e\ell E_z , \mu_2\}/\lambda_{SO}$ = \{(1,0.5), (1,-0.5), (1.5,0.5)\} 
        (solid blue, dashed red, and dash-dotted green, respectively) and $E_B = \lambda_{SO}$. For all curves, the distance between the plates is long enough so that Eq. \eqref{LongDistanceCasimir} is valid and $\Lambda, \Gamma = 0$.}
        \label{fig3}
    \end{center}
\end{figure}

In Fig. 3 we plot the long distance Casimir energy for two identical layers as a function of the chemical potential $\mu_1$ of one of the plates. 
In this situation, the Chern numbers of the monolayers do not always coincide, which opens up the possibility of changing the attractive or repulsive character of the quantized Casimir force. 
The sign of the force can be tuned by properly choosing $\mu_1$, while its magnitude could also be tailored by modifying $E_z$ and $\mu_2$. We emphasize that this fine control of the Casimir interaction
is only possible due to the energy level quantization arising from the introduction of a magnetic field. As pointed out in ref. \cite{woods2017}, without this agent switching the sign of the dispersive Casimir 
force from repulsion to attraction breaks its quantized behaviour. 
We also comment that for $|\mu_1| > E_B$ and $|\mu_2| < E_B$ the Casimir force is a full mix of photo-induced topology and quantum Hall physics since 
$C_1 = C_{1,\text{QH}}$ and $C_2 = C_{2,\text{ph}}$.

The accuracy of the long distance approximation is not the same in all regions of the phase space. In fact, if we look at the upper and bottom portions 
of the phase diagram in Fig 2c and compare it to the numerical values obtained from Eq. \eqref{LongDistanceCasimir}, it becomes evident that this 
approximation is not as good as it is for smaller values of $\mu$. In Fig. 4a we plot the difference between the exact Casimir energy, obtained by full 
numerical integration of the Lifshitz formula, and the approximated solution in the long distance limit for different values of $\mu$. All curves tend 
to zero for large distances, but it is evident that the Casimir energy converges faster to its asymptotic expression for smaller values of the chemical 
potential. The reason for this discrepance is due to the fact that the asymptotic solution is obtained by neglecting the contribution arising from the 
longitudinal conductivity. However, as we can see in the inset of Fig. 4a, for larger values of $\mu$ the longitudinal conductivity increases near 
$i\xi = 0$ and, in this case, needs to be accounted in the low frequency expansion for the distances considered. Fig. 4b shows how our result is affected 
by introducing a non-negligible dissipation $\Gamma \neq 0$ in the monolayers. In this situation, the Casimir energy becomes less repulsive, and can be 
attractive for sufficiently high values of dissipation. To understand this from an analytical point of view, we once again expand the integrand of Eq. 
\eqref{Lifshitz} for small values of $\xi$. Since both $\sigma_{xx}$ and $\sigma_{xy}$ are non zero at $\xi = 0$, the dominant contribution of the 
Casimir energy comes from the DC conductivity tensor. For sufficiently small values of the dissipation ($\hbar\Gamma << \lambda_{SO}$), 
we have $\sigma_{xx}(0) \approx \sigma_{xx}'(0)\Gamma$ and $\sigma_{xy}(0) \approx \frac{\alpha c}{2\pi} C$, where we took advantage of the similar 
role played by $\Gamma$ and $\xi$ in Eqs. \eqref{sigmaxx} and \eqref{sigmaxy}. By retaining only the longitudinal conductivity and expanding the 
reflection matrices for small $\xi$ we obtain
\begin{equation}
\frac{\Delta E^{(0)}}{E_g} = \frac{4\alpha\Gamma}{c}\frac{\mbox{log}(\sigma_{1,xx}'/\sigma_{2,xx}')}{(\sigma_{2,xx}')^{-1} - (\sigma_{1,xx}')^{-1}}. \label{DissipationCasimir}
\end{equation}
\begin{figure}
\begin{center}
\includegraphics[width=0.98\linewidth]{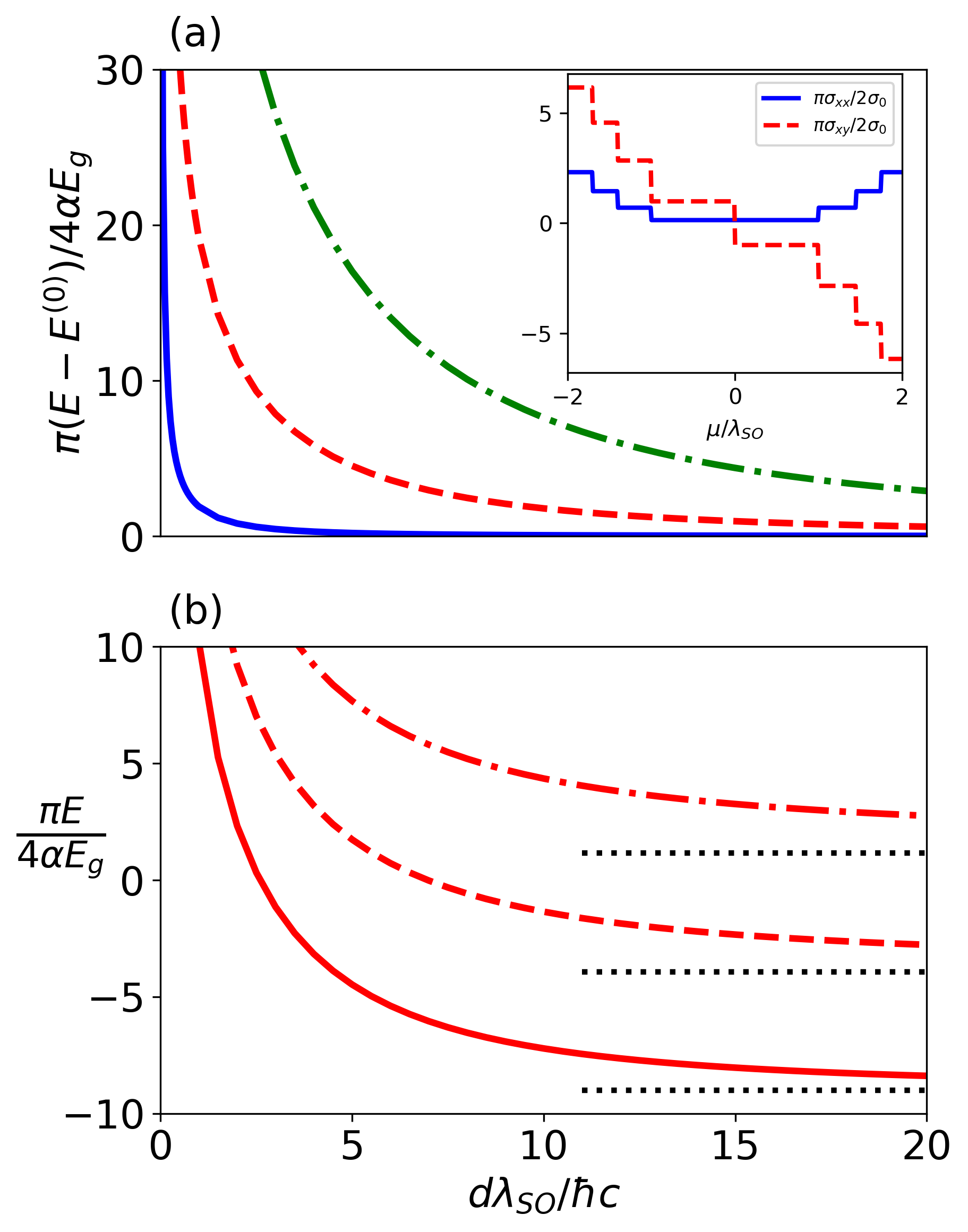}
\caption{(a) Long distance approximated expression given by Eq. \eqref{LongDistanceCasimir} subtracted from the Casimir energy as a function of distance. The chosen parameters are $\Lambda = 0$, $\Gamma = 0$, $e\ell E_z = \lambda_{SO}$, and $\mu = \{0.5,1.25,1.5\}\lambda_{SO}$ (solid blue, dashed red, and dash-dotted green lines, respectively). The inset shows the longitudinal and Hall conductivity at frequency $\hbar\xi = 0.01\lambda_{SO}$ as a function of the chemical potential. {(b)}. Casimir energy for $\Lambda = 0$, $e\ell E_z = \lambda_{SO}$, $\mu = 1.25\lambda_{SO}$, and $\hbar\Gamma = \{0,0.01,0.02\}\lambda_{SO}$ (solid, dashed, and dash-dotted lines, respectively) as a function of distance. The gray dotted lines are the approximated solutions in the long distance regime, given by the sum of Eqs. \eqref{LongDistanceCasimir} and \eqref{DissipationCasimir}.}
\label{fig4}
\end{center}
\end{figure}
For two identical layers, this expression simplifies to $\Delta E^{(0)}/E_g = 4\alpha\Gamma\sigma_{xx}'/c$. The dotted gray lines in Fig. 4b are given 
by the sum of eq. \eqref{LongDistanceCasimir} with eq. \eqref{DissipationCasimir}, and have a good agreement with the non approximated results (red curves).
\section{Finite-temperature Casimir energy}
\begin{figure}[h!]
\begin{center}
\includegraphics[width=0.98\linewidth]{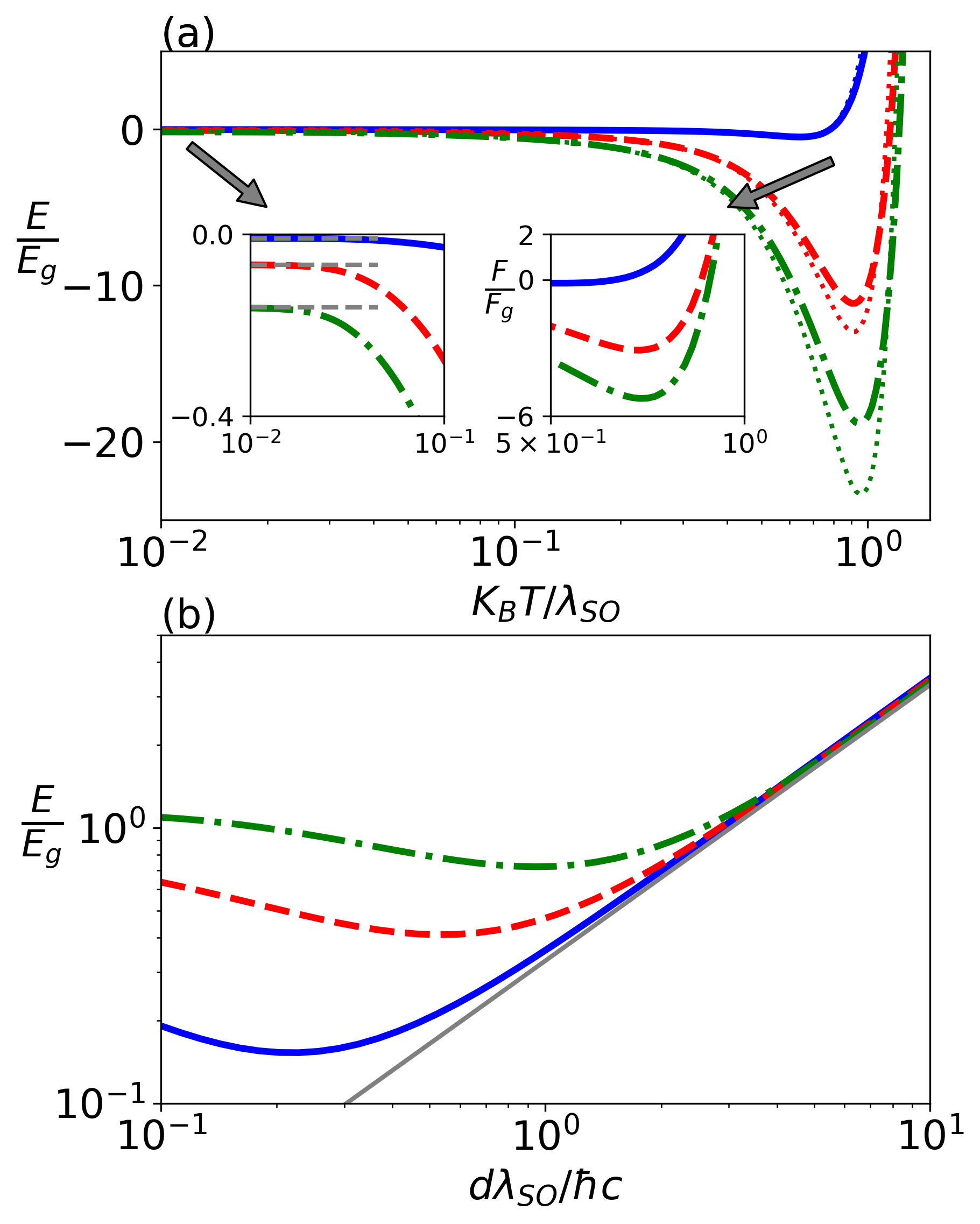}
\caption{{(a)} Dissipationless Casimir energy as a function of temperature for a separation distance of $d\lambda_{SO}/\hbar c = 10$. The dotted lines correspond to the approximated value of the $n = 0$ Matsubara frequency contribution presented in Eq. \eqref{HighTemperatureCasimir}. The left inset is a zoom of the plot, showing that each curve goes to its correspondent zero-temperature limit (dashed gray lines). The right inset shows the Casimir force in the region where it becomes attractive.\linebreak {(b)} Casimir energy as a function of distance for $k_BT = 10^{-3}\lambda_{SO}$ and $\hbar\Gamma = 0.01\lambda_{SO}$. The gray line corresponds to the $n = 0$ Matsubara contribution (Eq. \eqref{DissipationHighTemperatureCasimir}). In both plots, $\Lambda = 0$, $e\ell E_z = \lambda_{SO}$, and $\mu = \{0.5,1.25,1.5\}\lambda_{SO}$ (solid blue, dashed red, and dash-dotted green lines, respectively).}
\label{fig5}
\end{center}
\end{figure}

Now we consider thermal effects in the Casimir energy between two parallel graphene family layers. For finite temperatures the Casimir interaction energy 
follows from Eq. \eqref{Lifshitz} by considering the temperature dependent conductivity of the monolayers. The finite-temperature Casimir energy exhibits 
a rather distinct behaviour depending on whether or not we neglect dissipation in the materials. We first consider two identical dissipationless 
layers ($\Gamma = 0$). In Fig. 5a we show the normalized Casimir energy in the long distance regime as a function of temperature for different values 
of $\mu$.  For sufficiently small temperatures ($k_BT/\lambda_{SO} \lesssim 10^{-2}$), the Casimir energy is well described by the zero-temperature limit (left inset). 
As we increase the temperature, however, the Casimir force becomes more repulsive (right inset), reaching values two orders of magnitude greater than those at zero 
temperature in the region $k_BT/\lambda_{SO} \in [0.1,1]$ (below, near, and above room temperature for silicene, germanene, and stanene, respectively). 
This contrasts with the results found in the absence of a magnetic field, where even for smaller values of temperature ($k_BT/\lambda_{SO} \simeq 10^{-3}$, 
which corresponds to $T \sim 2,3K$ for plumbene and smaller for other materials) the force is attractive, reaching the zero-temperature limit only for 
$k_BT/\lambda_{SO} \simeq 10^{-4}$.\cite{woods2017}For $k_BT \simeq \lambda_{SO}$, the strength of the interaction reaches a maximum value and then 
rapidly becomes attractive due to thermal transitions between different Landau levels. 

This behaviour of the finite-temperature Casimir energy can be described analytically by considering only the contribution of the $n = 0$ Matsubara 
frequency in Eq. \eqref{Lifshitz}. Since for $\Gamma = 0$ and small values of $\xi$ we have $\sigma_{xx} = \sigma_{xx}'(0)\xi$, we get a contribution 
from both the longitudinal and Hall conductivities. By expanding the reflection matrices in $\xi$, taking the limit $\xi \rightarrow 0$, and keeping the 
lowest order contribution in $\alpha$, we get $\mathbb{R}_{sp} = \mathbb{R}_{ps} \approx 2\pi\sigma_{xy}/c$, $\mathbb{R}_{ss} \approx 0$, and 
$\mathbb{R}_{pp} \approx 2\pi\sigma_{xx}'k_\parallel$ (notice that the conductivities depend on temperature). Hence, we obtain
\begin{equation}
\frac{E}{E_g} = -\frac{16\pi^2 \sigma_{1,xy}\sigma_{2,xy}}{\hbar c \alpha}k_BTd + \frac{12\pi^2\sigma_{1,xx}'\sigma_{2,xx}'}{\hbar c \alpha}\frac{k_BT}{d}. \label{HighTemperatureCasimir}
\end{equation}
The dotted lines in Fig. 5a were obtained from the previous equation, and we note that there is a good agreement between this approximation and the 
numerical calculations. Eq. \eqref{HighTemperatureCasimir} highlights that the repulsive Casimir interaction arises from the Hall conductivity, 
which holds information about topology and quantum Hall physics, while the attractive counterpart is due to the longitudinal conductivity. 
In contrast to the zero-temperature limit, the positive and negative terms of eq. \eqref{HighTemperatureCasimir} have different distance scaling laws. 
As a consequence, the Casimir force becomes attractive at a temperature slightly lower than that associated to the situation $E = 0$, as shown in the 
right inset of Fig. 5a. In the regime where $k_BT \ll \lambda_{SO}$ but $T$ is high enough so that keeping only the $n = 0$ term is still a good 
approximation, we can discard the second term of eq. \eqref{HighTemperatureCasimir} and assume that $\sigma_{i,xy} \approx 2\sigma_0C_i/\pi$, which leads 
to $E/E_g = -(4\alpha c k_BTd/\hbar)C_1C_2$. This shows that applying a magnetic field on the materials increases the robustness of the topological 
features arising in the Casimir force, which are now present even at high temperatures ($k_BT \sim 0.1\lambda_{SO}$). On the left inset of Fig. 5a, 
the slope of each curve is proportional to the square of its corresponding Chern number, being higher for larger values of $\mu$.

The picture is not the same if we add dissipation to the system. In this case, $\sigma_{xx}(0) \neq 0$ and the limit $\xi \rightarrow 0$ of the 
reflection matrices yields $\mathbb{R}_{ss} = \mathbb{R}_{sp} = \mathbb{R}_{ps} = 0$ and $\mathbb{R}_{pp} = 1$. Hence, regardless of the topological 
phase of the materials, the $n = 0$ contribution to the Casimir energy is given by
\begin{equation}
\frac{E}{E_g} = \frac{2\zeta(3)}{\hbar c \alpha}k_BTd, \label{DissipationHighTemperatureCasimir}
\end{equation}
which is exactly the same result as the one found in the absence of a magnetic field. Note that there is no dependence on the conductivities and, 
more importantly, the result is the same for any non-zero value of $\Gamma$. Fig. 5b shows that, in this situation, even for temperatures of 
$k_BT/\lambda_{SO} = 10^{-3}$ the Casimir force is always attractive, as the case when $B = 0$. Hence, in order to see repulsion in a dissipative 
system one has to decrease the temperature to the sub-kelvin regime, which poses a serious difficult in measuring the Casimir force. 

This discrepance between the results obtained for a dissipationless and a dissipative system is no coincidence. Having a finite dissipation, even if arbitrarily small, 
is fundamentally different from having no dissipation. For metallic materials, this is equivalent to describe the system using a plasma ($\Gamma = 0$) or a Drude ($\Gamma \neq 0$) 
model, which leads to significantly different final results \cite{mostepanenko2015}. In principle, the Drude 
model is the most straightforward approach for taking into account the relaxation properties of conduction electrons, and should be applicable in the quasistatic limit 
(which corresponds to the high temperature regime of the Casimir force). In fact, a few experiments \cite{garcia2012,lamoreaux2011} show that the Casimir force is consistent with the 
calculations using this model. However, the plasma model approach, which usually appears in the literature in infrared optical phenomena, was shown to best 
describe the Casimir force between metals in most of the experiments \cite{decca2003,decca2007,banishev2013,decca2016}. Which model is the correct one to use when calculating the Casimir force is an open 
problem that still awaits its ending chapters. Here we provide a full description of the Casimir interaction between dissipative and dissipationless 
graphene family topological insulators and comment that the Drude versus plasma discussion can be extended to non-metallic systems. We emphasize that this extremely different behaviour of the 
dissipationless Casimir force in the graphene family does not show up when $B = 0$, where it was shown that both models differ only by a factor $2$ 
(as the case of metallic plates) \cite{woods2017}. Hence, the addition of a magnetic field to the problem allows one not only to avoid difficulties concerning temperature 
but also to clearly discriminate both descriptions and investigate fundamental questions in physics.
\section{Conclusions and final remarks}
In this work we presented a complete study of the Casimir effect between graphene family materials under the influence of an external magnetic field. 
We have shown that, for sufficiently large distances, the zero-temperature Casimir force between two monolayers is proportional to 
the product of their corresponding Chern numbers and can be made attractive or repulsive by properly doping the materials. 
The remarkable possibility of switching the sign of the Casimir force in the flatland is exclusively due to the interplay between photo-induced 
phase transitions and quantum Hall physics brought by the presence of a magnetic field. Due to the intrinsically two-dimensional nature of the graphene family materials,
in order to test this prediction in an experiment one should be able to measure the Casimir force in a parallel surface configuration.
In the last few years we have witnessed important advances regarding the alignment and stability in Casimir force measurement setups with parallel plates \cite{norte2018,sedmik2018,mostepanenko2020_2}, 
which may be extended to two-dimensional topological insulators in the near future. On the other hand, 
a metallic sphere close to the graphene family sheet could be used to explore the topological features of the system, and is a more adopted scheme for measuring the Casimir force
in laboratory \cite{neto2017}. Furthermore, we can use the 
chemical potential as a substitute of the circulary polarized laser field, which can be very useful to avoid increasing the materials temperature due to photo-absorption. 
We believe this may help experimental investigations of the Casimir force between graphene family materials near zero temperature. 

Finally, we analyzed the thermal effects on the Casimir energy, which proved to be significantly different from previous studies using the graphene family 
materials. We have shown that the dissipative model imply a rather trivial high temperature Casimir force, where no topological physics is present. 
However, the picture is very different if we assume a dissipationless system. In this case, we found repulsive forces of the order of the zero-temperature graphene-graphene interaction, 
even at room temperature, and the topological features are not completely lost. For instance, let us consider two stanene monolayers at 
$T = 300K$ and separated by a distance $d = 20\mu$m. By evaluating the Casimir force per unit area for $\Lambda = 0$ and $e\ell E_z = \lambda_{SO}$, 
we obtain $F \approx 0.1$nPa at $\mu = 1.25\lambda_{SO}$ and $F \approx 0.2$nPa at $\mu = 1.5\lambda_{SO}$, while the zero-temperature graphene-graphene Casimir 
pressure is given by $F_g \approx 0.05$nPa. Finally, since some experiments show that the Casimir 
force is in agreement with the plasma model and others with the Drude model, we emphasize the importance of our findings 
in the dissipationless limit. In fact, a huge difference between both situations allows drawing more confident conclusions when testing which model 
correctly describes the Casimir interaction in the extended graphene family.

\section{Acknowledgements}
The authors thank V. Henning and R. Guimar\~aes for enlightening discussions. Y.M. and. C.F. acknowledge funding by the Coordena\c c\~ ao de Aperfei\c coamento de Pessoal de N\'ivel Superior (CAPES). C.F. acknowledges  Conselho Nacional de Desenvolvimento Cient\'ifico e Tecnol\'ogico (CNPq) - research grant 310365/2018-0. W.K.-K. acknowledges the Laboratory Directed Research and Development program of Los Alamos National Laboratory for funding under project number 20190574ECR.
%
%\newpage
%%%%%%%%%%%%%%%%%%%%%%%%%%% Bibliografia %%%%%%%%%%%%%%%%%%%%%%%%%%%%%%%%

%%%%%%%%%%%%%%%%%%%%%%%%%%%%%%%%%%%%%%%%%%%%%%%%%%%%%%%%%%%%%%%%%%%%%%%%%
%


\begin{thebibliography}{99}

\bibitem{bordag2009} Bordag, M. {\em et al.} Advances in the Casimir effect. {\em OUP Oxford}  Vol. {\bf 145} (2009).

\bibitem{dalvit2011} Dalvit, D. {\em et al.} eds. Casimir physics. {\em Springer} Vol. {\bf 834} (2011).

\bibitem{buhman2013} Buhmann, S. Y. Dispersion Forces I: Macroscopic quantum electrodynamics and ground-state Casimir, Casimir-Polder and Van der Waals forces. {\em Springer} Vol. {\bf 247} (2013).

\bibitem{bordag2001} Bordag, M., Mohideen, U., Mostepanenko, V. M. New developments in the Casimir effect. {\em Physics reports} {\bf 353} 1-3 (2001).

\bibitem{woods2016} Woods, L. M. {\em et al.} A materials perspective on Casimir and van der Waals interactions. {\em Rev. Mod. Phys.} {\bf 88,} 045003 (2016).

\bibitem{farina2006} Farina, C. The casimir effect: some aspects. {\em Brazilian journal of physics} {\bf 36,} 1137-1149 (2006).
 
\bibitem{casimir1948} Casimir, H. B. G. On the attraction between two perfectly conducting plates. {\em Proc. Kon. Nederland Akad. Wetensch} {\bf 51, } 793-795 (1948).

\bibitem{parsegian2006} Parsegian, V. A. Van der Waals Forces: A Handbook for Biologists, Chemists, Engineers, and Physicists {\em Cambridge U.P.,
New York} (2006)

\bibitem{israelashivili2011} Israelashivili, J. Intermolecular and Surface Forces. {\em 3rd ed. Academic Press, New York} (2011).

\bibitem{distasio2012} DiStasio, R. A., Jr., von Lilienfeld, O. A., Tkatchenko, A. Collective many-body van der Waals interactions in molecular systems. {\em Proc. Natl. Acad. Sci. U. S. A.} {\bf 109,} 14791 (2012).

\bibitem{autumn2000} Autumn, K. {\em et al.} Adhesive force of a single gecko foot-hair. {\em Nature} {\bf 405,} 681 (2000).

\bibitem{autumn2002} Autumn, K. {\em et al.} Evidence for van der Waals Adhesion in Gecko Setae. {\em Proc. Natl. Acad. Sci. U. S. A.} {\bf 99,} 12252 (2002).

\bibitem{serry1998} Serry, F. M., Walliser, D., Maclay, G. J. The role of the Casimir effect in the static deflection and stiction of membrane strips in microelectromechanical systems (MEMS). {\em J. Appl. Phys.} {\bf 84,} 2501 (1998).

\bibitem{buks2001} Buks, E., Roukes, M. L. Stiction, adhesion energy, and the Casimir effect in micromechanical systems, {\em Phys. Rev. B} {\bf 63,} 033402 (2001).

\bibitem{broer2013} Broer, W. {\em et al.} Significance of the Casimir force and surface roughness for actuation dynamics of MEMS. {\em Phys. Rev. B} {\bf 87,} 125413 (2013).

\bibitem{sedighi2015} Sedighi, M., Palasantzas, G. Influence of low optical frequencies on actuation dynamics of microelectromechanical systems via Casimir forces. {\em J. Appl. Phys.} {\bf 117,} 144901 (2015).

\bibitem{rosa2018} Rodrigues, J. R., Gusso, A., Rosa, F. S. S., Almeida, V. R. Rigorous analysis of Casimir and van der Waals forces on a silicon nano-optomechanical device actuated by optical forces. {\em Nanoscale} {\bf 10,} 3945-3952 (2018).

\bibitem{mostepanenko2009} Klimchitskaya, G. L., Mohideen, U., Mostepanenko, V. M. The Casimir force between real materials: experiment and theory. {\em Rev. Mod. Phys.} {\bf 81, } 1827-1885 (2009).

\bibitem{johnson2011} Rodriguez, A. W., Capasso, F., Johnson, S. G. The Casimir effect in microstructured geometries. {\em Nat. Photon.} {\bf 5,} 211-221 (2011).

\bibitem{tang2017} Tang, L. {\em et al.} Measurement of non-monotonic Casimir forces between silicon nanostructures. {\em Nat. Photon.} {\bf 11,} 97-101 (2017).

\bibitem{grushin2011} Grushin, A. G., Cortijo, A. Tunable Casimir repulsion with three-dimensional topological insulators. {\em Phys. Rev. Lett} {\bf 106,} 020403 (2011).

\bibitem{cortijo2011} Grushin, A. G., Rodriguez-Lopez, P., Cortijo, A. Effect of finite temperature and uniaxial anisotropy on the Casimir effect with three-dimensional topological insulators, {\em Phys. Rev. B} {\bf 84,} 045119 (2011).

\bibitem{lopez2014} Rodriguez-Lopez, P., Grushin, A. G. Repulsive Casimir effect with Chern insulators. {\em Phys. Rev. Lett.} {\bf 112}, 056804 (2014).

\bibitem{wilson2015} Wilson, J. H., Allocca, A. A., Victor Galitski, Repulsive Casimir force between Weyl semimetals, {\em Phys. Rev. B} {\bf 91,} 235115 (2015).

\bibitem{ma2014} Ma, J., Zhao, Q., Meng, Y., Magnetically controllable Casimir force based on a superparamagnetic metametamaterial. {\em Phys. Rev. B} {\bf 89,} 075421 (2014)

\bibitem{santos2009} G\'omes-Santos, G. Thermal van der Waals interaction between graphene layers. {\em Phys. Rev. B} {\bf 80, } 245424 (2009).

\bibitem{woods2010} Drosdoff, D., Woods, L. M. Casimir forces and graphene sheets. {\em Phys. Rev. B} {\bf 82,} 155459 (2010).

\bibitem{sarabadani2011} Sarabadani, J. {\em et al.} Many-body effects in the van der Waals-Casimir interaction between graphene layers. {\em Phys. Rev. B} {\bf 84,} 155407 (2011) 

\bibitem{bordag2012} Bordag, M., Klimchitskaya, G. L., Mostepanenko, V. M. Thermal Casimir effect in the interaction of graphene with dielectrics and metals. {\em Phys. Rev. B} {\bf 86} 165429 (2012).

\bibitem{mostepanenko2013} Klimchitskaya, G. L., Mostepanenko, V. M. Van der Waals and Casimir interactions between two graphene sheets. {\em Phys. Rev. B} {\bf 87} 075439 (2013).

\bibitem{tsoi2014} Tsoi, S. {\em et al.} van der Waals screening by single-layer graphene and molybdenum disulfide. {\em ACS Nano} {\bf 8,} 12410-12417 (2014).

\bibitem{gobre2014} Gobre, V. V., Tkatchenko, A. Scaling laws for van der Waals interactions in nanostructured materials. {\em Nat. Commun.} {\bf 4,} 2341 (2014).

\bibitem{bimonte2017} Bimonte, G., Klimchitskaya, G. L., Mostepanenko, V. M. Thermal effect in the Casimir force for graphene and graphene-coated substrates: Impact of nonzero mass gap and chemical potential. {\em Phys. Rev. B} {\bf 96} 115430 (2017).

\bibitem{mostepanenko2020} Klimchitskaya, G. L., Mostepanenko, V. M. Quantum field theoretical description of the Casimir effect between two real graphene sheets and thermodynamics. {\em Phys. Rev. D} {\bf 102} 016006 (2020).

\bibitem{gusynin2005} Gusynin, V. P., Sharapov, S. G. Unconventional Integer Quantum Hall Effect in Graphene. {\em Phys. Rev. Lett.} {\bf 95,} 146801 (2005).

\bibitem{gusynin2007} Gusynin, V. P., Sharapov, S. G., Carbotte, J. P. Magneto-optical conductivity in graphene. {\em J. Phys.: Condens. Matter} {\bf 19,} 026222 (2007).

\bibitem{goerbig2011} Goerbig, M. O. Electronic Properties of Graphene in a Strong Magnetic Field. {\em Rev. Mod. Phys.} {\bf 83,} 1193 (2011).

\bibitem{nicol2013} Tabert, C. J., Nicol, E. J. Magneto-optical conductivity of silicene and other buckled honeycomb lattices {\em Phys. Rev. B} {\bf 88}, 085434 (2013).

\bibitem{macdonald2012} Tse, W. K., MacDonald, A. H. Quantized Casimir force. {\em Phys. Rev. Lett.} {\bf 109,} 236806 (2012).

\bibitem{tarik2014} Cysne, T. {\em et al.} Tuning the Casimir-Polder interaction via magneto-optical effects in graphene. {\em Phys. Rev. A} {\bf 90,} 052511 (2014).

\bibitem{ni2012} Ni, Z. {\em et al.} Tunable bandgap in Silicene and Germanene. {\em Nano Lett.} {\bf 12,} 113-118 (2012).

\bibitem{nicol2012} Stille, L., Tabert, C. J., Nicol, E. J. Optical signatures of the tunable band gap and valley-spin coupling in silicene. {\em Phys. Rev. B} {\bf 86,} 195405 (2012).

\bibitem{ezawa2013} Ezawa, M. Photoinduced Topological Phase Transition and a Single Dirac-Cone State in Silicene. {\em Phys. Rev. Lett.} {\bf 110,} 026603 (2013).

\bibitem{ezawa2015} Ezawa, M. Monolayer Topological Insulators: Silicene, Germanene and Stanene. {\em J. Phys. Soc. Jpn.} {\bf 84,} 121003 (2015).

\bibitem{kortkamp2017} Kort-Kamp, W.J. M. Topological phase transitions in the photonic spin Hall effect. {\em Phys. Rev. Lett.} {\bf 119,} 147401 (2017)

\bibitem{woods2017} Rodriguez-L\'opez, P., Kort-Kamp, W. J. M., Dalvit, D. A. R., Woods, L. M. Casimir force phase transitions in the graphene family. {\em Nat. Commun.} {\bf 8,} 14699 (2017).

\bibitem{reynaud2006} Lambrecht, A., Neto, P. A. M., Reynaud, S. The Casimir effect within scattering theory. {\em New Journal of Phys.} {\bf 8,}243 (2006).

\bibitem{dalvit2018} Ledwith, P., Kort-Kamp, W. J. M., Dalvit, D. A. R. Topological phase transitions and quantum Hall effect in the graphene family. {\em Phys. Rev. B} {\bf 97,} 165426 (2018).

\bibitem{xiang2017} Yu. X.-L., Huang, L., Wu, J. From a normal insulator to a topological insulator in plumbene. {\em Phys. Rev. B} {\bf 95,} 125113 (2017).

\bibitem{karch2010} Karch, J. {\em et al.} Dynamic Hall Effect Driven by Circularly Polarized Light in a Graphene Layer. {\em Phys. Rev. Lett.} {\bf 105, } 227402 (2010)

\bibitem{mostepanenko2015} Klimchitskaya, G. L., Mostepanenko, V. M. Casimir free energy of metallic films: discriminating between Drude and plasma model approaches. {\em Phys. Rev. A} {\bf 92, } 042109 (2015).

\bibitem{garcia2012} Garcia-Sanchez, D. {\em et al.} Casimir force and {\em in situ} potential measurements on nanomembranes. {\em Phys. Rev. Lett.} {\bf 109, } 027202 (2012).

\bibitem{lamoreaux2011} Sushkov, A. O. {\em et al.} Observation of the thermal Casimir force. {\em Nat. Phys.} {\bf 7, } 230-233 (2011)

\bibitem{decca2003} Decca, R. S. {\em et al.} Improved tests of extra-dimensional physics and thermal quantum field theory from new Casimir force measurements. {\em Phys. Rev. D} {\bf 68, } 116003 (2003)

\bibitem{decca2007} Decca, R. S. {\em et al.} Tests of new physics from precise measurements of the Casimir pressure between two gold-coated plates. {\em Phys. Rev. D} {\bf 75, } 077101 (2007).

\bibitem{banishev2013} Banishev, A. A. {\em et al.} Casimir interaction between two magnetic metals in comparison with nonmagnetic test bodies. {\em Phys. Rev. B} {\bf 88, } 155410 (2013)

\bibitem{decca2016} Bimonte, G., L\'opez, D., Decca, R. S. Isoelectronic determination of the thermal Casimir force. {\em Phys. Rev. B} {\bf 93, } 184434 (2016)

\bibitem{norte2018} Norte, R. A. {\em et al.} Platform for Measurements of the Casimir Force between Two Superconductors. {\em Phys. Rev. Lett} {\bf 121, } 030405 (2018)

\bibitem{sedmik2018} Sedmik, R., Brax, P. Status Report and first Light from Cannex: Casimir Force Measurements between flat parallel Plates. {\em J. Phys.: Conf. Ser.} {\bf 1138} 012014 (2018)

\bibitem{mostepanenko2020_2} Mostepanenko, V. M., Klimchitskaya, G. L. Recent measurements of the Casimir force: Comparison between experiment and theory {\em Mod. Phys. Lett. A} {\bf 35, } 2040007 (2020)

\bibitem{neto2017} Hartmann, M., Ingold, G.-L., Neto, P. A. M. Plasma versus Drude modeling of the Casimir force: beyond the proximity force approximation. {\em Phys. Rev. Lett.} {\bf 119, } 043901 (2017).

%\bibitem{grushin2014} Grushin, A. G., G\'omez-Le\'on, \'A., Neupert, T. Floquet fractional Chern insulators. {\em Phys. Rev. Lett.} {\bf 112,} 156801 (2014).

%\bibitem{gomez2014} G\'omez-Le\'on, \'A., Delplace, P., Platero, G. Engineering anomalous quantum Hall plateaus and antichiral states with ac fields. {\em Phys. Rev. B} {\bf 89,} 205408 (2014).

%\bibitem{nicol2013} Tabert, C. J., Nicol, E. J. Valley-spin polarization in the magneto-optical response of silicene and other similar 2D crystals. {\em Phys. Rev. Lett.} {\bf 110,} 197402 (2013).

%\bibitem{macdonald2010} Tse, W. K., MacDonald, A. H. Giant magneto-optical Kerr effect and universal Faraday effect in thin-film topological insulators. {\em Phys. Rev. Lett.} {\bf 105,} 057401 (2010).

%\bibitem{tabert2013} Tabert, C. J., Nicol, E. J. Magneto-optical conductivity of silicene and other buckled honeycomb lattices. {\em Phys. Rev. B} {\bf 88,} 085434 (2013).


\end{thebibliography}
\end{document}